\documentclass[10pt]{scrbook}
\usepackage[T1]{fontenc}  
\usepackage{lmodern}
\usepackage[ansinew]{inputenc}
\usepackage[german,english]{babel} 

\usepackage{geometry}
\addto\captionsgerman{

}

\usepackage{graphics,graphicx}
\usepackage{color}
\usepackage{longtable}
\usepackage{amsmath,exscale,tipa} 
\usepackage{amstext} 
\usepackage{wasysym,latexsym} 
\usepackage {subfigure}
\usepackage{threeparttable} 
\usepackage{breakcites} 
\usepackage{xfrac} 
\usepackage{yfonts}

\usepackage{scrlayer-scrpage}
\usepackage{booktabs}

\deffootnote{10pt}{0pt}{\thefootnotemark\ }

\usepackage{url}

\newcommand{\manuscript}{{\em Manuscript}}
\def\farcm{\hbox{$.\mkern-4mu^\prime$}}

\begin{document}


	\centerline{\includegraphics[width=12.0cm]{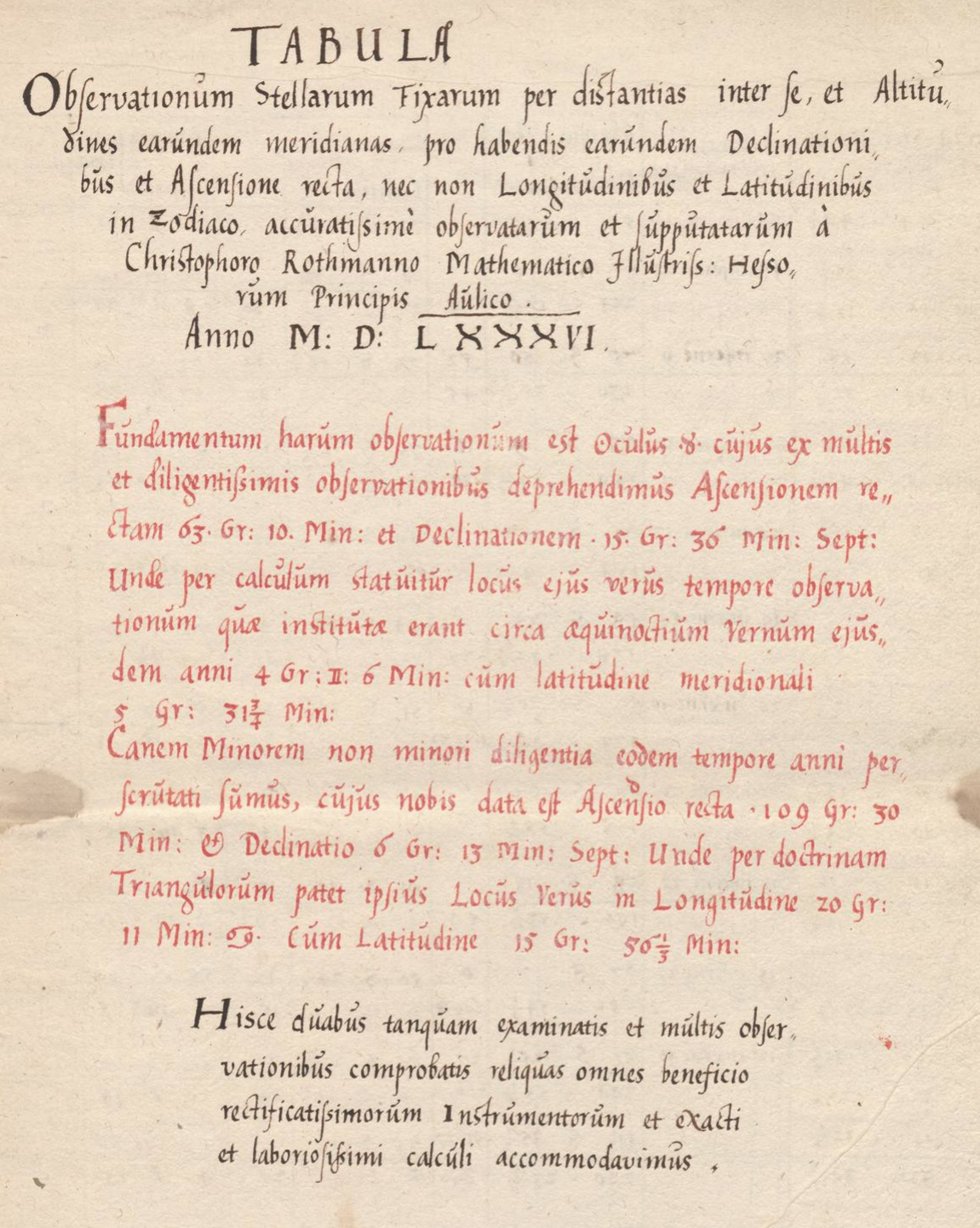}}





\newpage

\noindent Frontispiece: The first page of the catalogue of 121 stars in a Vienna manuscript of Rothmann. \"Osterreichische Nationalbibliothek (N$^o$ 80 e codice 10686)\footnote{\url{digital.onb.ac.at/RepViewer/viewer.faces?doc=DTL_6615943&order=1&view=SINGLE}}

\vspace*{0.5cm}

Translation:

\begin{quote}

Table of observations of fixed stars through the distances between them and
their meridional altitudes in order to obtain their declinations and right
ascension, as well as their ecliptic longitudes and latitudes, [the stars]
most accurately observed and computed by Christopher Rothmann,
court mathematician  of the Illustrious Sovereign of Hessen.
{In} the year 1586.
\vspace*{0.3cm}

The basis of these observations is the Eye of Taurus for which we obtained a right 
ascension  $63^\circ10'$ and declination $+15^\circ36'$ from many
very careful observations, from which by calculation its true location  at
$4^\circ6'$\,\gemini\ with  southern latitude $5^\circ31'45''$ is determined
for the epoch of the observations which were made near the vernal equinox 
of the same year.\newline
With not less diligence have we examined
at the same time of the year Canis Minor, whose right ascension
$109^\circ30'$ and declination $+6^\circ13'$ is given by us; from which by the
theory of triangles its true location appears in longitude  $20^\circ11'$\,\cancer\ with
latitude $15^\circ56'20''$.
\vspace*{0.3cm}

To these two thus examined and confirmed with many observations,
we have accommodated all the others with the benefit of 
instruments made most accurate and of exact and most laborious calculation.
\end{quote}

\chapter*{The star catalogue of Wilhelm IV, Landgraf von Hessen-Kassel\\[1cm]
  \Large{\textit{Andreas Schrimpf (Marburg) \& Frank Verbunt
      (Nijmegen, The Netherlands)}}}

\addcontentsline{toc}{chapter}{\thechapter{ The star catalogue of Wilhelm IV, Landgraf von Hessen-Kassel}\\ 
\textit{Andreas Schrimpf (Marburg) \& Frank Verbunt (Nijmegen, The Netherlands)}}

\markboth{G.\,Wolfschmidt: Applied and Computational Historical
  Astronomy}{A.\,Schrimpf et al.: The star catalogue von Wilhelm IV,
  Landgraf von Hessen-Kassel} \vspace{1.5\baselineskip}
\label{Schrimpf}
\index{Wilhelm IV [Landgraf von Hessen-Kassel from 1567
    to 1592] (1532 --1592)}

\subsection*{Abstract}

{\small Near the end of the 16$^{th}$ century Wilhelm IV, Landgraf von
  Hessen-Kassel, \index{Wilhelm IV [Landgraf von Hessen-Kassel from 1567
    to 1592] (1532--1592)} set up an observatory with the main goal to
  increase the accuracy of stellar positions primarily for {use in astrology and for calendar
  purposes}. A new star catalogue was compiled from measurements of altitudes
  and angles between stars and a print ready version was prepared
  listing measurements as well as equatorial and ecliptic coordinates
  of stellar positions. Unfortunately, this catalogue appeared in print
  not before 1666, long after the dissemination of Brahe's
  \index{Brahe, Tycho (1546--1601)} catalogue. {With the data given in the
  manuscript we are able to analyze the accuracy of measurements and computations.} {The measurements and the computations
  are very accurate, thanks to the instrument maker and mathematician
  Jost B\"urgi. \index{B\"urgi, Jost (1552--1632)}
The star catalogue is more accurate by a factor two
  than the later catalogue of Tycho Brahe.} \index{Brahe, Tycho (1546--1601)} \vspace{0.4cm}

{\selectlanguage{german}
\subsection*{Zusammenfassung}

{\small Zur Erh÷hung der Genauigkeit von Sternkoordinaten, die f³r die Berechnung von {Horoskopen und Kalendern} notwendig wurden,  gr³ndete und betrieb Wilhelm IV, \index{Wilhelm IV [Landgraf von Hessen-Kassel from 1567
    to 1592] (1532--1592)} Landgraf von Hessen--Kassel, im letzten Drittel des 16. Jahrhunderts eine Sternwarte in Kassel. Aus neuen Messungen von Sternh÷hen und -abstõnden wurde ein nahezu druckfertiger Sternkatalog zusammengestellt, in dem neben õquatorialen und ekliptischen Sternkoordinaten auch Messdaten aufgef³hrt sind. Ungl³cklicherweise erschien der Katalog als Kopie erst 1666, lange nach der Verbreitung von Tycho Brahes Katalog.
{Anhand der Messdaten und Koordinaten im Katalog sind wir dazu in der Lage, die Genauigkeit der damaligen Messungen und Berechnungen zu analysieren.} Beides, die Messungen und die Berechnungen, sind auch dank der Zuarbeit des Instrumentenbauers und Mathematikers Jost B³rgi sehr prõzise. Im Endergebnis ist der Katalog etwa um einen Faktor 2 genauer als der spõtere Katalog von Tycho Brahe.} 
} 

\section{Introduction}

At the end of Chapter\,XIX of his \textit{Astronomia Nova} Kepler \index{Kepler, Johannes (1571 -- 1630)} (1609)
\nocite{kepler09}
explains that an error of 8 arcminutes remains when he compares the
best model in the method of Ptolemaios \index{Ptolemaios, Klaudios (100 -- 170)} -- which only uses motions in
circles -- with {with several observations of Mars made} by Brahe. He then states, {we translate from Latin}:
\begin{quote}
 Well, if I had judged that 8 minutes of longitude could be ignored,
 I would have sufficiently corrected the hypothesis (i.e. the
 bisection of the eccentricity) of Chapter\,XVI.  Now because they
 could not be ignored, {just these 8 minutes} led the way to the reformation of all
 astronomy, and were made the subject matter of a large part of this
 work.
\end{quote}
which includes the discovery that a planet moves in an ellipse with the Sun at
one focal point. The measurement accuracy of 2 arcminutes
achieved by Tycho Brahe was instrumental in Kepler's revolutionary
discovery.  Kepler was well aware who showed Brahe the means of
accurate observing: in his booklet  {\em  De stella tertii honoris in Cygno 
Narratio Astronomica} on the nova of 1604 in Cygnus he had written 
(Kepler 1606, edition by Frisch 1859, p.769,  {we translate from Latin}):
\nocite{kepler06}\nocite{frisch59}
\begin{quote}
Jost B\"urgi, \index{B\"urgi, Jost (1552--1632)} the maker of
moving models, who, although he knows no
languages, yet in the science and consideration of mathematics
easily surpasses many professors of these. He possesses a dexterity
so peculiar to him that a later age may celebrate him as a
true leader in this genre, no less than D\"urer \index{D³rer, Albrecht (1471 -- 1528)} 
in painting, whose fame grows imperceptibly like a tree with time.
\newline
\ldots\newline
Previously in the service of the most illustrious Landgraf von Hessen Wilhelm
(whose industriousness and diligence in the celestial science were bigger
than one would look for in a sovereign and whose very famous findings
stimulated Tycho Brahe to emulate him).\newline
\ldots\newline
It was in that time that Rothmann, the astronomer of the Landgraf, 
especially devoted himself to his work on the fixed stars.
\end{quote}

Indeed, Brahe visited the observatory of Wilhelm\,IV in 1575 and from
1585 Wilhelm\,IV and Rothmann corresponded with Brahe on
methods of improving measurement accuracy (Hamel
2002).\nocite{hamel02} Such methods were implemented in the
observatory in Kassel, and in this article we show that the accuracy
achieved by Wilhelm\,IV and his collaborators Jost B\"urgi \index{B\"urgi, Jost (1552--1632)} and
Christoph Rothmann surpassed that of Brahe. The manuscript star
catalogue was produced by Rothmann in 1587, and predates the larger
star catalogue of Tycho Brahe by a decade. (The Brahe catalogue was
circulated as a manuscript in 1598, and printed for the first time  in 1602.)
\nocite{brahe98}\nocite{brahe02}\nocite{dreyer15}\nocite{dreyer16}

The manuscript star catalogue became widely known after its publication by Curtz \index{Curtz, Albrecht (1471 -- 1528)}
(1666)\nocite{curtz66} and its importance was acknowledged by its
inclusion in the star catalogues of Hevelius \index{Hevelius, Johannes (1611 -- 1687)} (1690)\nocite{hevelius90}
and Flamsteed \index{Flamsteed, John (1646 -- 1719) } (1725).\nocite{flamsteed25} Hamel (2002) remarks that
the star catalogues published by Copernicus \index{Copernicus, Nicolaus (1473 -- 1543)}
(1543)\nocite{copernicus43} in the {\em De revolutionibus} and
Reinhold \index{Reinhold, Erasmus (1511 -- 1553)} (1551)\nocite{reinhold51} in the {\em Prutenicae Tabulae}
were still the original catalogue by Ptolemaios, from about 150\,CE,
merely corrected for precession. This highlights the crucial
contribution of Wilhelm\,IV in being the first to alert the European
astronomical world to the need of new observations for more accurate
positions of the stars and -- since the positions of planets were
measured with respect to nearby stars -- of the planets, {\em and to
  act on it}\,! In the course of time the appreciation of the
importance of Wilhelm\,IV gradually moved to the background, as
attention focused on Tycho Brahe.  It is therefore welcome that
J\"urgen Hamel (2002), inspired by articles of Rudolf Wolf \index{Wolf, Rudolf (1816 -- 1893)}
(1878) and Johann Adolf Repsold \index{Repsold, Johann Adolf (1838 -- 1919)} (1919), brought it back to our
attention. {Hamel's } monograph describes the astronomical research in
Kassel under Wilhelm\,IV, and includes an analysis by Eckehard
Rothenberg, of the Archenhold Observatory in Berlin, of the accuracy
of the star catalogue of Wilhem\,IV.

We have made a detailed analysis of the star catalogue of Wilhelm\,IV
and of the measurements given in it, {where } we provide a machine-readable
version of the catalogue (Verbunt \&\ Schrimpf
2021). \nocite{verbuntschrimpf21} In this article we 
describe the historical context and give 
more details about the instruments.  We
present the extant version of the star catalogue of Wilhelm\,IV in
Section\,\ref{s:manuscript}, the accuracy of the underlying
measurements in Section\,\ref{s:measurements}, the accuracy of the
computations made for it in Section\,\ref{s:computations}, and the
accuracy of the star positions in the catalogue in
Section\,\ref{s:positions}. Our conclusions are summarized in
Section\,\ref{s:conclusions}, together with an outlook on further
research. First, however, it is appropriate to describe astronomy at
Kassel.

\section{Astronomy at Kassel}

Two paintings from 1577 by Kaspar van der Borcht \index{van der Borcht, Kaspar (1576 -- 1610) } show Landgraf
Wilhelm\,IV at age 45, and his wife Landgr\"afin Sabina von
W\"urttemberg. \index{Sabina von
W\"urttemberg [Landgrõfin] (1549 -- 1581) } The paintings, which form a unit, are now in the
{Gem\"aldegalerie} Alte Meister in Kassel\footnote{\url{https://altemeister.museum-kassel.de/69429/37796/0/147/b1/1/0/objekt.html}}, and are described by
Kirchvogel (1967),\nocite{kirchvogel67} as follows. 
Wilhelm and Sabina are standing on a platform of the castle at Kassel.
On the left behind Wilhelm we see the top of armature surrounding a
celestial globe, on the right behind him two men hold a sextant.  One
of the men is Eberhard Baldewein, \index{Baldewein, Eberhard (1525 -- 1593)} head of the workshop of Wilhelm\,IV;
the other man may well be Tycho Brahe.  To the left of the
Landgr\"afin two astronomical instruments are depicted: a torquetum on
a high stand and an azimuthal quadrant.  In the distance the valley of
the river Fulda is visible with a smaller castle and geometric
gardens. Of the instruments, only the globe still exists, in the same
museum as the diptich. The images of constellations on the globe are a
later addition (Gaulke 2007).\nocite{gaulke07}

With an azimuthal quadrant the azimuth and altitude of a star can be
measured. A torquetum is used to make measurements directly in
altazimuthal, equatorial or ecliptic coordinates. The armature of a
globe can serve to convert a celestial position of a star into
altazimuthal coordinates or vice versa. For accurate results it is
better to use a sextant or quadrant for the measurements, and to use
spherical geometry equations for coordinate conversions.
\vspace*{1.cm}

Hamel (2002) describes the astronomy in Kassel.  Wilhelm\,IV was born
in 1532, and educated privately, and in the year 1546/7 at the
Gymnasium in Strassbourg.  His first known astronomical observations
in Kassel date from 1558, and include measurements with the torquetum
of the comet of that year.  In 1560 Wilhelm founded the observatory at
the castle, and observations there were made by him personally, and by
assistants. The frequency of observations by Wilhelm\,IV personally
necessarily became less after he succeeded his father Philipp \index{Philipp [Landgraf von Hessen from 1509/1518  to 1567] (1504 -- 1567)} as
landgrave of Hessen in 1567.  

Astrology relies on accurate positions of the planets, which rely on
accurate positions of the stars. Good calendars require good knowledge
of the motion of the Sun.  The large discrepancies between different
editions of the star catalogue of Ptolemaios, due to the build-up of
errors in transmission over many centuries, convinced Wilhelm that
astrology and calendar-making could only be saved by new observations,
as accurate as possible.  Observations with the torquetum were
replaced in 1565 with observations with a quadrant and a sextant, and
coordinate transformations were done with the equations of spherical
geometry rather than with the globe.  The clockmaker Eberhard
Baldewein made among others a 5-foot wooden quadrant, an azimuthal
quadrant, and an {armillary sphere} of messing. Observations were made a.o.\
of the comets of 1577, 1580 and 1585, and the New Star of
1572. Wilhelm's observations of the New Star were admired throughout
Europe for their accuracy.  They showed that the new star was further
away than the Moon, and Wilhelm understood that this invalidated the
division of the Universe by Aristoteles into a sublunar region full of
change and a never-changing supralunar region. His letters on
the subject show that he believed in astrology. In 1575 Brahe visited {Wilhelm\,IV}
to discuss the means of accurate observing.

In 1579 Jost B\"urgi \index{B\"urgi, Jost (1552--1632)} was hired as an instrumentmaker, and set to work
building more accurate clocks and quadrants. {A detailed biography of B³rgi is given by Staudacher (4th edition, 2018).} Around 1580 Christoph
Rothmann joined the observatory as its main astronomer and
mathematician.  In 1584 Paul Wittich, who had studied with Brahe for
four months, visited Kassel and informed the astronomers there about
the transversal lines that Brahe had introduced in the scales of his
instruments. B\"urgi incorporated this in his improvements
(see Figure\,\ref{f:nonius}).  279 stellar altitudes were
measured with the quadrant between 19 January and 27 March of 1585,
timed with the precision clock constructed by B\"urgi. Later the
altitudes were measured at the northern and southern meridian
passages. From these measurements stellar positions were
determined. From 22 February 1585 angles between stars were measured
as an alternative method to determine stellar positions. An analysis
by Peters \&\ Sawitsch (1849)\nocite{peters49} shows that measurements
on the orbit of the comet of 1585 were accurate to one arcminute. At
the end of 1585 Rothmann found that measurements could be repeated to
an accuracy of one third of an arcminute. A steady exchange of letters
with Brahe ensued, continued from 1585 until 1590. Brahe
published this exchange in 1596, in a volume of 340 pages; interest in
these letters was such that Brahe reprinted them in 1601 and 1610
(Hamel 2002).

\begin{figure}
\centerline{\includegraphics[width=12.cm]{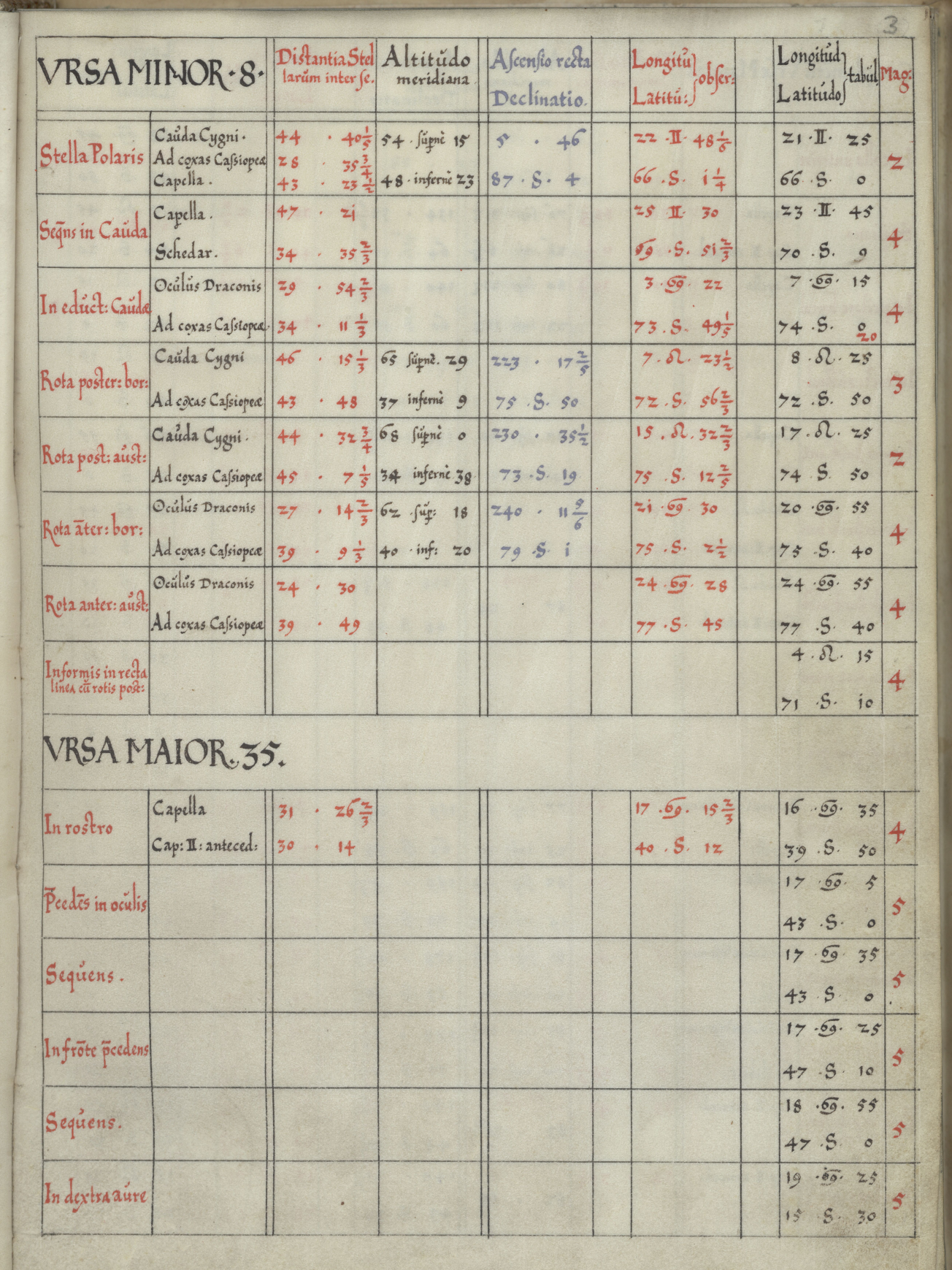}}
\caption{The first page of \manuscript\ (ORKA, see footnote 3) \label{f:firstpage}}
\end{figure}

The stellar positions thus obtained were collected in various preliminary
catalogues of 23, 58, and 121 entries. Finally a catalogue with new positions
for 387 stars was readied for publication by Rothmann in 1587; the 
manuscript  looks almost finished but was not published at the time. The
first publication of the catalogue appeared {in} 1666 as a part of the ''Historia coelestis'' by Albert Curtz (1666).

In 1589 Rothmann described \textit{Observations of the fixed
    stars} in a manuscript, now in the collection of the University of
  Kassel, which has made it available on its online platform Open
  Repository Kassel (ORKA), with its shelf mark \mbox{$2^\circ$
  Ms. astron. 5, no.7}\footnote{
  \url{orka.bibliothek.uni-kassel.de/viewer/image/1350031335734/364/}}.
The Latin text of this handbook of astronomy has been edited
  by Granada et al. (2003), with an introduction in
  German.\nocite{granada03} Wolf (1878)\nocite{wolf78} gives a chapter by chapter
  summary of the contents of the Handbook.  The manuscript of the star
  catalogue of Wilhelm\,IV, hereafter \manuscript, is also in the
  collection of the University of Kassel, available on ORKA, with
  shelf mark \mbox{$2^\circ$ Ms. astron. 7}\footnote{
  \url{orka.bibliothek.uni-kassel.de/viewer/image/1336543085355/1/}}.

\section{The star catalogue of Wilhelm\,IV\label{s:manuscript}}

The first page of the manuscript of the Star Catalogue of
Wilhelm\,{IV}, shown in Figure\,\ref{f:firstpage}, starts immediately
with the catalogue, without introduction.  On 71 pages \manuscript\
contains in the 7th and 8th column the ecliptic coordinates and
magnitudes of 1032 entries: the 1028 stars in the star catalogue in
the Almagest of Ptolemaios, and four added stars. The ecliptic
longitudes from the Almagest have been increased by $21^\circ15'$ to
correct for precession between the epoch of that catalogue and
1586. \manuscript\ follows the order of the constellations in the
Almagest, but the names or descriptions of the stars, in the first
column, and the order in which stars are listed within a constellation
can be different.  Thus in Ursa Minor the `End of the tail' and the
`rectangle' in Ptolemaios are the Pole Star and front wheel and back
wheel (rota anterior and posterior) in \manuscript.  $\zeta$ and
$\eta$ are listed before $\beta$ and $\gamma$ in the Almagest, but
after them in the \manuscript\ (Figures\, {\ref{f:firstpage} and} \ref{f:uma}).

\begin{figure}
\centerline{\includegraphics[width=8.cm]{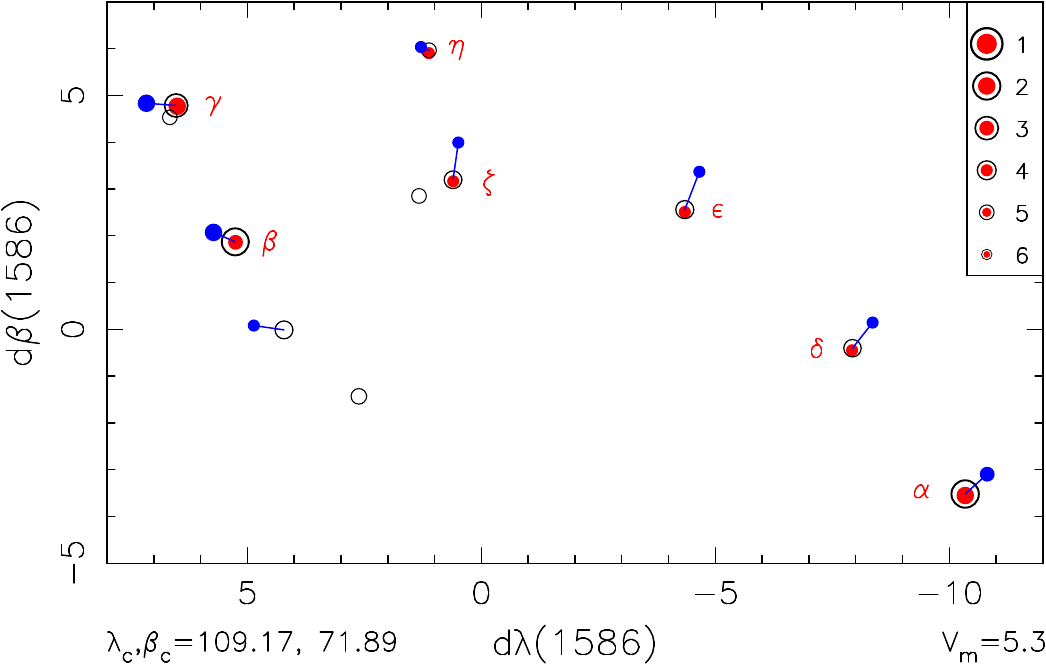}}
\caption{ Comparison of the positions of stars in \manuscript\ (red) with
those for stars with $V<5.3$ computed for epoch 1586 from the modern
HIPPARCOS-2 catalogue (van Leeuwen 2007, open circles), and of the
star catalogue in the Almagest converted to 1586 (blue). The scales
are in degrees with respect to the projection center indicated left
below. The inset gives the magnitude scale.\label{f:uma}}
\end{figure}

For 388 entries the third column gives measurements of the angular
distances to stars mentioned in the second column, and for all but one
of these the ecliptic coordinates {resulting from the observations in Kassel} are given in the sixth column. For
346 entries \manuscript\ in addition gives measurements of the
altitude at meridional passage in the fourth column and equatorial
coordinates in the fifth column.  All angles and coordinates are in
degrees, (arc)minutes and fractions of (arc)minutes; in the case of
ecliptic longitudes the angle is given within the constellation
indicated: 22\,\gemini\,48$\frac{1}{6}$ is $22^\circ48'10''$ within
Gemini, i.e. $\lambda=82^\circ48'10''$.  Northern and southern
latitudes or declinations are indicated with S and M, respectively.
Northern meridional altitudes may be measured above or below the pole
star and are indicated {\em superne} and {\em inferne}, respectively.

Figure\,\ref{f:uma} illustrates that the positions in \manuscript\ are
much more accurate than those in the Almagest, and indeed sufficiently
accurate to enable unambiguous identification. The three stars that are
(knowingly) repeated in the Almagest occur twice also in \manuscript.
The first entry in Perseus is identified with the double star cluster
h\&$\chi$\,Persei, the first entry in Cancer with the star cluster Praesepe.
\manuscript\ thus contains 384 independent entries, 382 stars and
two clusters of stars.

The absence of an introduction to the catalogue in \manuscript\ means
that its epoch is not given. To determine it, we follow Wolf (1878) in
turning to a manuscript by Rothmann, a catalogue of 121 stars
including a preamble, which he wrote in 1586 and which appears to be an
intermediate report of his work on the star catalogue. A copy of this
manuscript must have been sent to Brahe, and later was passed to
Kepler with other papers from Brahe. It is now part of the collection
of the \"Osterreichische National Bibliothek in Wien (N$^o$ 80 e
codice 10686). The layout of this catalogue by Rothmann is very
similar to that of \manuscript, and the epoch of the catalogue is
given as 1586 in a preamble, which we reproduce with a translation as
the frontispiece (see also Wolf 1878,
p.129.\nocite{wolf78} Wolf still had access to the original version in
Kassel).  Comparison of the right ascensions of the stars in this
catalogue with those in the manuscript in Kassel show that with few
exceptions they are identical.  From this we agree with Wolf
  (1878) in concluding that the epoch of the larger star catalogue is
1586.

Curtz (1666) gives equinox 1593 for the catalogue.  It appears likely
that he adapted the equinox as an approximate correction for the offset
of $6'$, found by Brahe, in the right ascensions.  As seen in the
preamble of the Vienna catalogue, the right ascensions are based on
the position of Aldebaran, and to a lesser extent of Procyon, which
are used as primary fundamental stars.  Computing the positions of
these stars from modern data in the HIPPARCOS-2 catalogue (van Leeuwen
2007)\nocite{leeuwen07}, we confirm this offset, which is also evident
in the right ascensions of the catalogue as a whole (Verbunt \&\
Schrimpf 2021).\nocite{verbuntschrimpf21} 
Wolf (1878, p.131) {shows that this offset may be explained} as due to the wrong parallax
assumed for the Sun.

\section{Accuracy of the measurements\label{s:measurements}}

Important innovations of the star catalogue in \manuscript\ are
inclusion of the measurements on which the positions of the stars are
based, and inclusion of both ecliptic and equatorial positions. 
This permits the reader to check the coordinates. It enables
us to directly quantify the accuracy of the measurements.

\begin{figure}
\centerline{\includegraphics[width=9.cm]{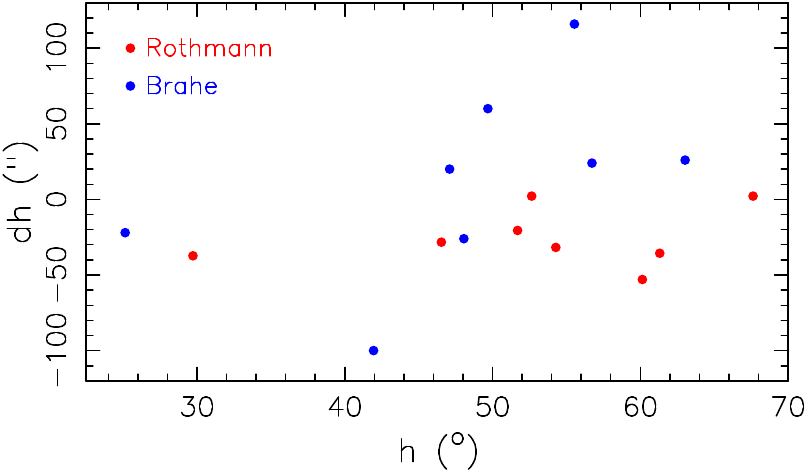}}
\caption{Errors in altitude measurements by Rothmann, computed
with Eq.\,\ref{e:altfit}, compared to those with the mural quadrant by
Brahe for his reference stars (Wesley 1978). Altitudes in  Brahe's observatory 
Hven are about $4^\circ36'$ lower than those in Kassel.\label{f:refstars}}
\end{figure}

With $h$ the true altitude, $h_\mathrm{a}$ the apparent altitude and $R$ the
refraction by the atmosphere, we have for meridian passage in the south at
a location at geographic latitude $\phi_\mathrm{G}$:
\begin{equation}
h_\mathrm{a} = h+ R = {90^\circ}-\phi_\mathrm{G} +\delta +R
\label{e:refrac}\end{equation}
We correct the equatorial coordinates from the HIPPARCOS catalogue for each entry 
for proper motion and precession to obtain $\delta$ in 1586, 
and apply Eq.\,\ref{e:refrac} with an approximate equation for
$R$ given by S\ae mundsson (1986) with $\phi_\mathrm{G}=51^\circ18'48''$
for Kassel to compute the apparent altitude in 1586 $h_\mathrm{a,{HIP}}$.
The question then arises {whether the values in {\manuscript\ } refer to the apparent altitude $h_\mathrm{a}$ or to the true altitude $h$} . To avoid this ambiguity, we at first
limit the comparison to altitudes $h\geq29^\circ$, where according to Rothmann
the refraction is zero, and thus $h=h_\mathrm{a}$. A fit to these altitudes measured
at southern meridian passage gives
\begin{equation}
dh \equiv h - h_\mathrm{a,{HIP}} = -1\farcm8(2) + 0\farcm023(3)\,h(^\circ);\qquad h\geq29^\circ ,
\label{e:altfit}\end{equation}
{where the numbers between brackets give the one-sigma error in the preceding decimal.}
The measurement error $dh$ is smallest at the zenith, and increases
with distance to the zenith.   A fit to the values at all altitudes $h>0$, that
includes the altitudes for northern superior and inferior meridian
passages (computed with appropriate adaption of Eq.\,\ref{e:refrac}),
we find  compatible within the errors to Eq.\,\ref{e:altfit}, {\em
  provided we assume that the listed values are the apparent
  altitude {$h_\mathrm{a}$}}. The spread around the fit to $dh$ is
$\sigma=0\farcm74$.  A fit to the same data, but correcting the \manuscript\
values for refraction $R_\mathrm{R}$ according to Rothmann, gives an
incompatible fit:
\begin{equation}
\Delta h \equiv h+R_\mathrm{R} - h_\mathrm{a,{HIP}} = -1\farcm0(1) + 0\farcm011(2)\,h(^\circ)
\label{e:altfitb}\end{equation}
This supports our finding that all tabulated altitudes refer to the {apparent} altitude $h_\mathrm{a}$.

\begin{figure}
\centerline{\includegraphics[width=9.cm]{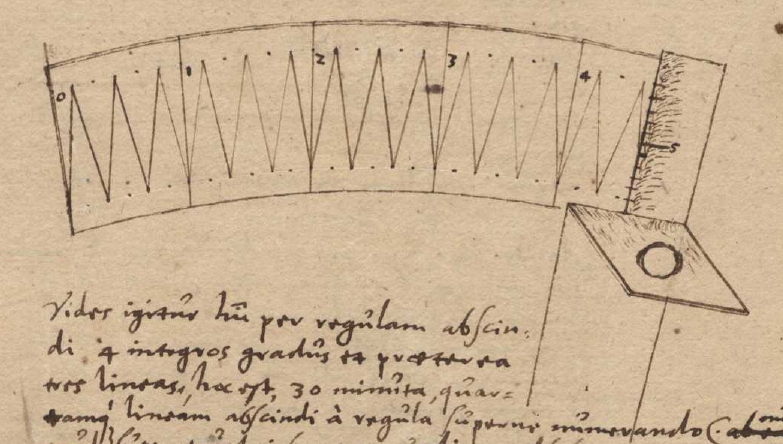}}
\caption{The use of transversal lines to improve reading accuracy. 
 B\"urgi makes each transversal line cover $10'$, and adds a vertical scale
to the alidade with $1'$ spacing. The alidade indicates $4^\circ35\farcm6$.
From the Handbook by Rothmann, p.13\,r (ORKA, see footnote 2).)\label{f:nonius}}
\end{figure}

Wesley (1978)\nocite{wesley78} investigated the measurement accuracy of
Brahe and gives numbers for altitude measurements with the large
mural quadrant for eight of Brahe's reference stars.  In Figure\,\ref{f:refstars} 
we compare these with the altitude errors for the same stars as measured by
Rothmann. It is seen that the measurements in Kassel are more accurate,
on average by a factor two (Verbunt \&\ Schrimpf 2021).

The other measurements are those of angles between stars, of which 500
are given in \manuscript. These too we compare with the angles computed from
HIPPARCOS data, after correction for proper motion. The accuracy of the
measurements in \manuscript\ is $\sigma=0\farcm73$, virtually the same 
as those for the altitude measurements. There is no systematic offset, and no
dependence of the measurement accuracy on the angle (Verbunt \& Schrimpf 2021).

Hamel (2002) remarks astutely that Brahe and Wilhelm\,IV went
different ways in their pursuit of accuracy.  Brahe increased the size
of his instruments, culminating in his large mural quadrant. The
clockmaker B\"urgi applied his genius to meticulous metalworking on a
small scale. From our results above we may conclude that the approach
in Kassel gave better results. As an example we reproduce in
Figure\,\ref{f:nonius} the drawing by Rothmann of the nonius of the
sextant made by B\"urgi. This may be compared with the illustration by
Brahe (1602b, p.123 in the edition of Dreyer 1823).\nocite{braheb02}\nocite{dreyer23}

\section{Accuracy of the computations\label{s:computations}}

The presence of B\"urgi in Kassel leads one to expect that the computations are
done with high accuracy. Independently from Napier, \index{Napier, John (1550 -- 1617)} and probably before
him around 1580, B\"urgi invented the logarithm, or perhaps more accurately
the antilogarithm. To replace a multiplication of $a$ and $b$ with an addition,
the standard way is to use a table of logarithms, whereas B\"urgi used exponents with base $B$:
\begin{equation}
\mathrm{standard}:\quad\log ab = \log a + \log b;\qquad \mathrm{B\"urgi}:
\quad B^aB^b = B^{a+b}
\end{equation}
B\"urgi also invented a completely new method to rapidly compute the
sine function to high accuracy, and in 1592 presented a manuscript to
emperor Rudolf\,II \index{Rudolf\,II [Holy Roman Emperor from 1576 to 1612] (1552 -- 1612)} in Prague with a sine table for every minute from
$0^\circ$ to $90^\circ$, i.e.\ 5400 values with 5 to 7 sexagesimal
places (Folkerts et al.\ 2016).\nocite{folkerts16} Thus B\"urgi could
efficiently make the large number of computations required for the
catalogue.

The most straightforward computation from measurement to coordinate
is the conversion of the altitude at southern or northern meridian passage
into declination. For southern altitudes this can be done with Eq.\,\ref{e:refrac},
for northern inferior and superior altitudes with adapted versions of that
equation. As a first step, we compare listed altitudes with catalogued
declinations for all entries at $h\geq29^\circ$, where refraction, according to Rothmann,
is zero. With only 4 exceptions, the computation is exact for $\phi_\mathrm{G}=51^\circ19'$. 
Thus we establish that this is the geographic latitude used in the computations.
Next we  compare the altitudes and declinations for $h<29^\circ$. For
all but one southern meridian altitude the conversion of altitude to declination
is exact, with $\phi_\mathrm{G}=51^\circ19'$ as above, if we do not
apply a correction for refraction, i.e.\ {\em if the listed altitudes 
are the true altitudes {$h$}}. 
{This contradicts the result that we obtained from
comparison of the listed altitudes with the measurements.} We have
no explanation for this discrepancy: it appears to be a genuine error
in \manuscript. Surprisingly, the inferior northern culminations agree
better, but not exactly, with the declinations given in manuscript
if they are {considered not as true, but {apparent} altitudes in the computations!} For  351 of the 378 listed altitudes
the conversion to declination in \manuscript\ is exact. Of the 27 altitudes
which do not give an exact match, 22 are northern inferior altitudes.

More involved is the computation of the right ascension $\alpha$ from
its declination $\delta$ and its angular distance $\phi$ to a reference star
with known equatorial coordinates $\alpha_\mathrm{r}$,$\delta_\mathrm{r}$:
\begin{equation}
\alpha=\alpha_\mathrm{r} + \arccos\left(
\frac{\cos\phi-\sin\delta\sin\delta_\mathrm{r}}{\cos\delta\cos\delta_\mathrm{r}}\right)
\label{e:alpha}\end{equation} 
\manuscript\ contains 500 measured values of $\phi$, of which 410
independent values between entries for which \manuscript\ also gives
equatorial coordinates. For these values we can compare the right
ascension $\alpha_\mathrm{eq}$ computed with Eq.\,\ref{e:alpha} with
the value $\alpha_\mathrm{W}$ listed in \manuscript. The median value
of the difference $|\alpha_\mathrm{eq}-\alpha_\mathrm{W}|$ is $1.3''$,
the average and rms of $\alpha_\mathrm{eq}-\alpha_\mathrm{W}$ are
$0.7''$ and $6.6''$, respectively. Compared to the measurement errors
these errors are negligible. 

To check whether it is possible to obtain
all catalogued right ascensions starting with Aldebaran as only
reference star, we first compute the right ascensions of 21 stars for
which the measured angular distance to Aldebaran is given. For seven
of these angular distances to other stars are given, and right
ascensions for these other stars can now be computed. After 8
iterations 64 stars were used as reference stars, and right ascensions
found for all but one of the 343 entries with equatorial coordinates
listed in \manuscript.  The number of (arc)seconds in the catalogue
are indicated as fractions of (arc)minutes, and only 14 possible fractions
$F$ are used, such that $60F$ is always integer. Our calculations
indicate that at each iteration the catalogued value of
$\alpha_\mathrm{W}$, rather than the exact computed value, is used in the
next iteration.  The discretisation of the numbers for (arc)seconds
has no significant effect on the average and rms of
$\alpha_\mathrm{W}$.

The next computation we check is the conversion of equatorial into
ecliptic coordinates. The angle between the position in ecliptic 
coordinates computed from the equatorial coordinates in \manuscript\ 
using the standard equations and the position listed in \manuscript\ may
be computed with
\begin{equation}
\phi_\mathrm{eq-W} =
2\arcsin\sqrt{\sin^2\frac{\beta_\mathrm{eq}-\beta_\mathrm{W}}{2}+
\cos\beta_\mathrm{eq}\cos\beta_\mathrm{W}\sin^2\frac{\lambda_\mathrm{eq}-\lambda_\mathrm{W}}{2}}
\label{e:havercomp}\end{equation}
where subscripts eq and W indicate the computed and listed coordinates, respectively.
The median, average and rms of $\phi_\mathrm{eq-W}$ are $3.0''$, $4.1''$ and $4.2''$.
The conversion between ecliptic and equatorial coordinates thus is very accurate,
its errors negligible with respect to the measurement errors.

Ecliptic coordinates are listed for 41 entries that have no listed equatorial coordinates.
This may indicate that the equatorial coordinates of these entries were known, even
if not listed. Alternatively the positions $\lambda$,$\beta$ of these entries could be found
iteratively from their angular distances to two other entries with
known ecliptic coordinates $\lambda_i$,$\beta_i$, by solving the equation pair:
\begin{equation}
F_i(\lambda,\beta) = \cos\beta\cos\beta_i\cos(\lambda-\lambda_i)+\sin\beta\sin\beta_i
-\cos\phi_i = 0;\qquad i=1,2
\label{e:twophi}\end{equation}
In all 37 cases where we can check this, the sum $|F_1|+|F_2|$ is less than $10^{-4}$.

This leaves 4 entries for which the measurements given in \manuscript\ are insufficient
for the computation of ecliptic coordinates. The measurements that were used 
for these were not entered in \manuscript.

\section{Accuracy of the Catalogue\label{s:positions}}

\begin{figure}
\centerline{\includegraphics[width=10.cm]{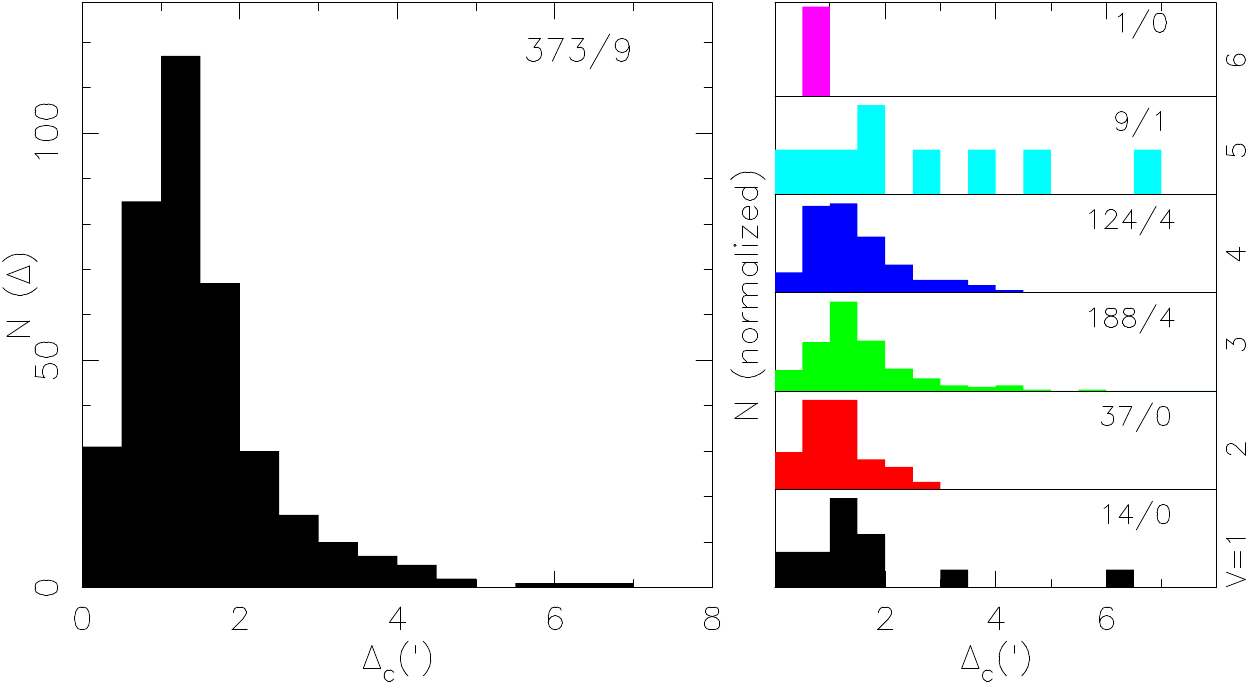}}

\caption{Positional error distribution, corrected for the 6' offset in $\alpha$, for the 382 entries in \manuscript,
after excluding two star clusters and three repeated entries, for all
magnitudes, and for each magnitude separately. The numbers in each
frame indicate the number of entries with errors smaller / larger than
the frame limit of $8'$.
\label{f:delta}}
\end{figure}

Magnitudes are listed for all 1032 entries in \manuscript, including
those with only the ecliptic coordinates from the Almagest of
Ptolemaios, corrected for precession. The star clusters Praesepe and
h\&$\chi$\,Per are listed as nebulous in both catalogues. To see whether the
magnitudes are copied from the Almagest or newly determined, we use
the HIPPARCOS identifications in our paper and in Verbunt \&\ van Gent
(2012)\nocite{verbunt12} to match the entries with new
measurements to entries in the star catalogue of Ptolemaios.  For
four entries a small shift in position leads to a different HIPPARCOS
identification, but Ptolemaios and the observer in Kassel may well
have looked at the same star.  For two entries that do not match, the
positions differ by several degrees, and the observed star was almost
certainly different.  Two of the entries in \manuscript,
$\epsilon$\,Aql (W\,999)\footnote{The entries of the machine--readable catalogue are numbered ''W xxx'' in order of their appearence, see Verbunt \&\ Schrimpf
2021.} and RR\,UMi (W\,39), are correctly marked as
not present in Ptolemaios.  For the Pleiades the situation is
confused, because the positions given in \manuscript\ differ
significantly from those in the Almagest after correction for
precession. As a result Atlas (W\,233) is indicated in \manuscript\ as
not present in the Almagest, whereas in fact it is. For 300 matches
the magnitude in \manuscript\ is identical to that in the
Almagest. For ten systems no match is found, because a different
position leads to a different HIPARCOS identification.  For the
remaining 72 entries the magnitude in \manuscript\ is brighter by two
for $\epsilon$\,Cep (W\,51), by one for 53, by slightly less than one
for 14, and fainter for 3 entries. $\lambda$\,Ori, marked nebulous in
the Almagest, is given magnitude 4 in \manuscript.  This indicates that
the magnitudes were estimated anew in Kassel. 

All entries in \manuscript\ can be identified with a HIPPARCOS
counterpart, indicative of a high accuracy. In Figure\,\ref{f:delta}
we show the distribution of the position differences
$\Delta_\mathrm{c}$ between the entry, corrected for the $6'$ offset
in $\alpha$, and its HIPPARCOS counterpart. Only 2.4\,\%\ has a
positional error larger than $8'$; the rms of the remaining ones is
$1'$. For comparison, in the star catalogue of Brahe (1598, 1602)
15\,\%\ has positional error larger than $10'$, many of them much
larger, and the rms of those with $\Delta<10'$ is $2'$ (Verbunt \&\
van Gent 2012).\nocite{verbunt12} Also if we limit the comparison
between \manuscript\ and Brahe to stars present in both, which
eliminates the fainter stars in Brahe, the conclusion holds that
\manuscript\ is a factor two more accurate (Verbunt \&\ Schrimpf
2021).  When no correction is made for the $6'$ offset, the rms error
of the positions in \manuscript\ is $1\farcm7$; still smaller than in
Brahe, and still with a much smaller fraction of very wrong positions.

\section{Conclusions and outlook}\label{s:conclusions}

Wilhelm IV was the first astronomer in Europe to start a systematic
observing program to obtain more accurate positions of the stars,
triggered by the need of accurate positions of the solar system bodies
for calendar purposes.  Thanks to the excellent instrument maker and
mathematician B\"urgi, the accuracy of the resulting catalogue is
significantly better than that of the later catalogue by Brahe.

Two more achievements of the astronomical activities at Kassel are
worth to point out: The superior clocks of Jost B\"urgi made it
possible to use time differences instead of angles, thus introducing a
new technique for determining distances of stars and paving the way to
transit measurements. Brahe doubted that this could be successful at
that time, but he was proved wrong by B\"urgi. Later in the
17$^{th}$ century  Olaf R{\oe}mer\index{R{\oe}mer, Olaf (1644 --
    1710)} re-established this method, which remained in use at
  groundbased observatories until replaced by photography. Position
  determination based on timing is still used in a very sophisticated
  way by modern astrometric satellites like HIPPARCOS and
  Gaia. Second, to our knowledge \manuscript\ is the first star
catalogue to list equatorial coordinates.

Copies of the catalogue as well as astronomical manuscripts from
Kassel were distributed in Europe at the beginning of the 17$^{th}$
century, the observations at Kassel are mentioned in quite a few
historical documents (Hamel 2002).

\vspace{1ex} There is still some work to be done. The documents at
Kassel contain more lists of stellar positions than the four versions
of the catalogue mentioned by Hamel (2002). We are in progress of
comparing these intermediate catalogues with the four already known
(but not yet fully analysed) versions in order to understand the
improvements of measurements and their reduction. {We also intend to
have a further look at the magnitudes listed in \manuscript\ for all
1032 entries.} We intend to present these investigations in a future
paper.

\section{Acknowledgement}
We thank Dr.\,Frederik Bakker (Center for the History of Philosophy
and Science, Radboud University) for his help with the translation of the Latin texts.

\section{Bibliography}

{\small 
\begin{description}
\setlength{\itemsep}{0pt}
\item
\textsc{Brahe, Tycho}.
\newblock \emph{Manuscript Star Catalogue}.
\newblock 1598, see Dreyer (1916).

\item
\textsc{Brahe, Tycho}.
\newblock \emph{Astronomiae Instauratae Progymnasmata}.
\newblock Uraniborg, Hven and Prague, 1602, see Dreyer (1915).

\item
\textsc{Brahe, Tycho}.
\newblock \emph{Astronomiae Instauratae Mechanica}.
\newblock Levinus Hulsius, N{\"u}rnberg, 1602b, see Dreyer (1923).

\item
\textsc{Copernicus, Nicolaus}.
\newblock \emph{De revolutionibus orbium coelestium}.
\newblock Petreius, N{\"u}rnberg, 1543.

\item
\textsc{Curtz, Albert}.
\newblock \emph{Historia Coelestis}.
\newblock Utzschneider, Augsburg, 1666.

\item
\textsc{Dreyer, Johan Ludvig~Emil}.
\newblock \emph{Tychonis Brahe Dani Scripta Astronomia, Vol.II}.
\newblock Gyldendal, Copenhagen, 1915.

\item
\textsc{Dreyer, Johan Ludvig~Emil}.
\newblock \emph{Tychonis Brahe Dani Scripta Astronomia, Vol.III}.
\newblock Gyldendal, Copenhagen, 1916.

\item
\textsc{Dreyer, Johan Ludvig~Emil}.
\newblock \emph{Tychonis Brahe Dani Scripta Astronomia, Vol.V}.
\newblock Gyldendal, Copenhagen, 1923.

\item
\textsc{Flamsteed, John}.
\newblock \emph{Historia Coelestis Brittanicae, Vol.\,3}.
\newblock H. Meere, London, 1725.

\item
\textsc{Folkerts, Menso, Launert, Dieter, and Thom, Andreas}.
\newblock \emph{Jost {B\"u}rgi's method for calculating sines}.
\newblock \emph{Historia Mathematica}, 43, 133--147, 2016.

\item
\textsc{Frisch, Ch.~(ed.)}.
\newblock \emph{Joannis Kepleri astronomi Opera Omnia, Vol.II}.
\newblock Heyder \&\ Zimmer, Frankfurt a.M., 1859.

\item
\textsc{Gaulke, Karsten}.
\newblock \emph{Der Ptolem{\"a}us von Kassel. Landgraf Wilhelm IV von
  Hessen-Kassel und die Astronomie}.
\newblock Kassel, 2007.

\item
\textsc{Granada, Miguel~A., Hamel, J{\"u}rgen, and von Mackensen, Ludolf}.
\newblock \emph{Christoph {R}othmanns {H}andbuch der {A}stronomie von 1589}.
\newblock Harri Deutsch, Frankfurt a.M., 2003.

\item
\textsc{Hamel, J{\"u}rgen}.
\newblock \emph{Die astronomischen Forschungen in Kassel unter Wilhelm IV}.
\newblock Harri Deutsch, Frankfurt a.M., 2002.

\item
\textsc{Hevelius, Johannes}.
\newblock \emph{Catalogus stellarum fixarum}.
\newblock Johann Stolle, Danzig [Gda\'nsk], 1690.

\item
\textsc{Kepler, Johannes}.
\newblock \emph{De stella tertii honoris in Cygno Narratio Astronomica}.
\newblock Paulus Sessius, Prague, 1606 (see `Frisch 1859).

\item
\textsc{Kepler, Johannes}.
\newblock \emph{Astronomia Nova}.
\newblock Voegelin, Heidelberg, 1609.

\item
\textsc{Kirchvogel, Paul~Adolf}.
\newblock \emph{Wilhelm {IV}, {T}ycho {B}rahe, and {E}berhard {B}aldewein: the
  missing instruments of the {K}assel {O}bservatory}.
\newblock \emph{Vistas in Astronomy}, 9 (1), 109--121, 1967.

\item
\textsc{Peters, Christian August~Friedrich and Sawitsch,
  Alexei~Nikolajewitsch}.
\newblock \emph{Bestimmung der {B}ahn des {C}ometen von 1585}.
\newblock \emph{Astronomische Nachrichten}, 29, 209--224, 1849.

\item
\textsc{Reinhold, Erasmus}.
\newblock \emph{Prutenicae Tabulae coelestium motuum}.
\newblock Morhardus, T{\"u}bingen, 1551.

\item
\textsc{Repsold, Johann Adolf}.
\newblock \emph{Zur {G}eschichte der astronomischen {W}erkzeuge}.
\newblock \emph{Astronomische Nachrichten}, 209, 193--210, 1919.

\item
{
\textsc{Staudacher, Fritz}.
\newblock \emph{Jost B³rgi, Kepler und der Kaiser}.
\newblock  NZZ Libro, 4th edition 2018.
}

\item
\textsc{van Leeuwen, Floor}.
\newblock \emph{Validation of the new {HIPPARCOS} reduction}.
\newblock \emph{Astronomy and Astrophysics}, 474, 653--664, 2007.

\item
\textsc{Verbunt, Frank and van Gent, Robert~Harry}.
\newblock \emph{The star catalogues of {P}tolemaios and {U}lugh {B}eg}.
\newblock \emph{Astronomy and Astrophysics}, 544, A31:1--34, 2012.

\item
\textsc{Verbunt, Frank and Schrimpf, Andreas}.
\newblock \emph{The star catalogue of {W}ilhelm {IV}, {L}andgraf von
  {H}essen-{K}assel}.
\newblock \emph{Astronomy and Astrophysics}, in press, 1--11, 2021.

\item
\textsc{Wesley, Walter~G.}
\newblock \emph{The accuracy of {T}ycho {B}rahe's instruments}.
\newblock \emph{Journal for the History of Astronomy}, 9, 42--53, 1978.

\item
\textsc{Wolf, Rudolf}.
\newblock \emph{Die hessischen {S}ternverzeichnisse}.
\newblock \emph{Astronomische Mittheilungen der Eidgen{\"o}ssischen Sternwarte
  Z{\"u}rich}, 5, 125--164, 1878.

\end{description}}




\refstepcounter{chapter}

\newpage
\section*{Authors}

\section*{Prof.\,Dr.\ Andreas Schrimpf (Marburg)}

{\small Born in 1958 in Fulda, Hessen; 1976--1983 student at Philipps-Universit\"at Marburg, 
1986 PhD in solid state physics, supervisor Prof.\,Dr.\ Hans Ackermann in Marburg; 1987--1988 fellow 
of the Deutsche Forschungsgemeinschaft at University of Virginia, USA; 1994 habilitation in solid 
state physics at Philipps-Universit\"at Marburg; 1995/1996 visiting professorship at Universit\"at Kassel; 
starting in 1996 research staff member at Physics Department, Philipps-Universit\"at Marburg; starting 
2004 head of workgroup on History of Astronomy and Observational Astronomy at Physics Department, 
Philipps-Universit\"at Marburg; 2015 associate professor at Physics Department, Philipps-Universit\"at Marburg.\\
  Publications in solid state physics, history of astronomy, stellar astrophysics\\
  Recent projects: history of astronomy, focus on Hessen, Germany;
  observation and analysis of variable stars, focus on automated
  analysis of the Sonneberg Plate Archive.
\vspace{0.24cm} 

\noindent \emph{Philipps-Universit\"at Marburg, Fachbereich Physik, \\
Renthof 5, 35032 Marburg, Germany} \\ 
E-Mail: \url{andreas.schrimpf@physik.uni-marburg.de}.
}

\section*{Em.Prof.\,Dr.\ Frank Verbunt (Nijmegen)}

{\small Born in 1953 in Goirle, Noord-Brabant, Niederlande. 1971-1977 student at University of Utrecht,
1982 PhD in astronomy, supervisors Henk van Bueren (Utrecht) and Ed van den Heuvel (University of
Amsterdam), 1982-1984 postdoctoral researcher at the Institute of Astronomy, Cambridge (U.K.) and
1984-1989 at the Max Planck Institut f\"ur extraterrestrische Physik in Garching
bei M\"unchen; 1989-2011 Full Professor of High-energy Astrophysics at the University of Utreecht and
2012-2019  at Radboud University Nijmegen. From 2020 emeritus professor of Radboud University.\\
Publications in binary evolution, high-energy astrophysics, history of astronomy.\\
Recent projects: velocities of new-born neutron stars, accuracy of old star catalogues, the discovery
of proper motion.\\
\vspace{0.24cm} 

\noindent 
\emph{Department of Astronomy / IMAPP, Radboud University, PO Box 9500, 6500 GL Nijmegen, The Netherlands}\\
E-Mail: \url{f.verbunt@astro.ru.nl}.
}

\end{document}